\documentclass[journal]{IEEEtran}
\usepackage{times}
\usepackage{epsfig}
\usepackage{graphicx}
\usepackage{amsmath}
\usepackage{amssymb}
\usepackage{multirow,hhline,tabularx,url}

\begin{document}
\title{Accurate Alignment Inspection System for Low-resolution Automotive and Mobility LiDAR}

\author{
    \IEEEauthorblockN{Seontake Oh, Ji-Hwan You}
    \IEEEauthorblockA{School of Mechanical and Control Engineering, Handong Global University, Rep. Korea\\}
    \and
    \IEEEauthorblockN{Azim Eskandarian}
    \IEEEauthorblockA{Dept. of Mechanical Engineering\\
    Virginia Tech, VA, USA}
    \and
    \IEEEauthorblockN{ }
    \IEEEauthorblockA{ \\}
    \and
    \IEEEauthorblockN{Young-Keun Kim}
    \IEEEauthorblockA{School of Mechanical and Control Engin.\\
    Handong Global University, Rep. Korea \\
    ykkim@handong.edu\\
    }
}

\maketitle

\begin{abstract}
A misalignment of LiDAR as low as a few degrees could cause a significant error in obstacle detection and mapping that could cause safety and quality issues.
In this paper, an accurate inspection system is proposed for estimating a LiDAR alignment error after sensor attachment on a mobility system such as a vehicle or robot.
The proposed method uses only a single target board at the fixed position to estimate the three orientations (roll, tilt, and yaw) and the horizontal position of the LiDAR attachment with sub-degree and millimeter level accuracy.
After the proposed preprocessing steps, the feature beam points that are the closest to each target corner are extracted and used to calculate the sensor attachment pose with respect to the target board frame using a nonlinear optimization method and with a low computational cost.
The performance of the proposed method is evaluated using a test bench that can control the reference yaw and horizontal translation of LiDAR within ranges of 3 degrees and 30 millimeters, respectively. The experimental results for a low-resolution 16 channel LiDAR (Velodyne VLP-16) confirmed that misalignment could be estimated with accuracy within 0.2 degrees and 4 mm.
The high accuracy and simplicity of the proposed system make it practical for large-scale industrial applications such as automobile or robot manufacturing process that inspects the sensor attachment for the safety quality control.
\end{abstract}





\section{Introduction}
\IEEEPARstart{L}{iDAR} (light detection and ranging) is an essential perception sensor, along with camera and radar, to provide 3D data of their surroundings for smart mobility systems such as vehicles, drones, and robots.
Also, it has become an icon for representing autonomous vehicles and high-level vehicle ADASs (advanced driver-assistance systems).

LiDAR perceives the surroundings with a point cloud of 3D data that is used to generate a high-resolution 3D map and detect and recognize objects such as vehicles, pedestrians, and other obstacles.
For vehicles, one or more LiDAR systems can track the relative movements of the surrounding vehicles and other obstacles simultaneously to predict and avoid collisions and can detect road lanes and curbs. For robots or drones, it is generally used to map surroundings and avoid obstacles.

Although LiDAR is a powerful sensor for providing 3D data for a safe ADAS and autonomous system, it has not been implemented widely, especially on commercial vehicles, due to its high cost and bulkiness.
With the production of low-resolution LiDAR and the recent advent of technology such as solid-state LiDAR, the overall dimensions and the costs of mobility LiDAR have become affordable.
If the application of LiDAR expands on automobiles or robots for mass production, an important safety issue regarding the sensor misalignment can arise.

Since LiDAR measures the surroundings with respect to its coordinate frame, it is essential that sensors are installed with proper alignment for accurate and safe measurements.
Thus, an alignment error as low as a few degrees could cause a significant offset in detecting the position of an obstacle that could cause a catastrophic accident.
A sensor misalignment could occur during the assembly process, and LiDAR alignment should be inspected precisely after attachment for safety and quality control.
Therefore a simple but accurate inspection system should be designed to be easily applicable to the manufacturing and servicing of mobility systems.

Studies on methods to detect LiDAR attachment errors has drawn little attention in the research and industrial community, as described in Section II.
Almost all literature related to LiDAR alignment has focused on the external calibration of LiDAR with other sensors, such as cameras and multiple LiDARs 
\cite{fremont2008extrinsic, alismail2012automatic, geiger2012automatic, pusztai2017accurate}.
These studies require utilizing other sensors and multiple chessboard targets or a target at different poses, which may not be practical for applications in the automotive or mobility industries.

Point cloud registration is a widely used method to estimate the relative pose of 3D points from reference 3D data that minimizes the cost function of alignment errors \cite{moosmann2011velodyne, segal2009generalized}. However, these algorithms have a significant calculation load for a large amount of 3D data are needed to be processed iteratively. Since a short inspection time and low computational cost are preferred in the industrial applications, the alignment estimation algorithm should be lightweight and use minimal beam points.

Therefore, this paper proposes a new accurate and lightweight method to estimate the misalignment of a LiDAR on a mobility system body frame using a rectangular target board at one fixed pose.
The feature points of LiDAR beams, which are the beams closest to each target corner, are extracted after the proposed preprocessing of the point cloud data. Then, a nonlinear optimization is applied to solve for the orientation and translation of the LiDAR body frame with respect to the target board, which is assumed to have a predetermined pose from the mobility body frame.

The contributions of this paper can be summarized as follows. First, this system requires only one target board at a fixed pose without utilizing other sensors, which has practical industrial applications.
Second, the system can estimate the sensor alignment within a high level of sub-degree and sub-centimeter accuracy, even for a low-resolution LiDAR.
Last, the algorithm used for estimating the alignment pose is lightweight with low computation cost, for it uses only four feature 3D points instead of the whole point cloud data iteratively.

The accuracy and precision of the proposed system are evaluated with simulations and experiments using a target board and a designed test-bench. The results validated that the proposed algorithm can estimate the misalignment with sub-degree and millimeter accuracy in real-time for a low-resolution LiDAR with 16 vertical channels.

The paper is organized as follows. In Section II, related literature is reviewed. In Section III, the algorithm for estimating the sensor alignment is described in detail. In Section III, simulations are conducted to evaluate the proposed algorithm, which is followed by experiments in Section IV to validate the overall accuracy and precision of the alignment estimation.

\section{Related Work }
LiDAR alignment refers to either the internal calibration of sensor parts or the external geometric calibration with other sensors or objects.

The internal calibration of LiDAR optical parts, such as laser diodes and lenses, is the process to minimize optical distortions and errors of laser beam sight. The calibration is generally conducted by the sensor manufacturer at the factory level. A typical method detects single or multiple targets at a distance to estimate the internal parameters, including the pose of each laser diode, which minimizes the overall measurement errors \cite{muhammad2010calibration, bergelt2017improving}. This paper assumes that LiDAR is calibrated internally before the attachment.

Most literature related to LiDAR alignment and calibration falls in the category of external calibration of the LiDAR with other sensors, usually cameras.
The extrinsic calibration estimates the geometric pose of a camera body frame with respect to the LiDAR body frame using one or more checkerboard targets.
Different algorithms have been proposed to find the 3D-2D correspondences between the laser beam points and the image pixels on several shapes of target boards \cite{fremont2008extrinsic, alismail2012automatic, kim2019extrinsic, geiger2012automatic, pusztai2017accurate}.
A widely used target is a planar chessboard, and the correspondences of the plane and lines of the board are extracted with at least three different poses \cite { zhang2004extrinsic, verma2019automatic} or at least one pose \cite {zhou2018automatic}. Other target shapes, such as circular \cite{fremont2008extrinsic, alismail2012automatic}, trihedral \cite{gong20133d} and 3D cubic \cite{pusztai2017accurate}, have been introduced.
These previous works required multiple target surfaces or a target at different poses for higher accuracy, which may not be practical in the automotive or mobility manufacturing process.
In contrast, this paper uses only one planar target at one fixed pose to estimate the pose between the LiDAR and the target frame.

Additionally, the relative alignment of a LiDAR from a reference 3D object can be estimated by using a point cloud registration. Minimizing the errors between the measured point cloud and the reference points can be derived by using ICP (Iterative Closed Point) algorithms \cite{ moosmann2011velodyne,wang2014simulation}, generalized ICP algorithms \cite{segal2009generalized, heide2018calibration}, and NDT (Normal Distributions Transform) algorithms \cite{zaganidis2017semantic}. However, these methods require processing a large amount of point cloud data iteratively, which introduces a significant calculation load. There are studies that attempt to accelerated ICP for LiDAR processing such as using a variant of nearest neighbor and local search \cite{choi2012fast} and
weighted parallel iterative closed point with interpolation \cite{wang2018single}. 
Our paper efficiently extracts only a few 3D points to estimate the alignment that makes a lightweight algorithm to be realized on a typical industry computer.

There are several studies that have estimated the pose of an object using sparse beam points. The three-beam detection system was proposed to estimate the 6D pose of a dynamic object using three laser beam points and a camera \cite{kim2012developing, kim2014design, kim2015portable}. That system was composed of three laser distance sensors with 640 nm beam spots, which were detectable by an RGB camera and were shown to achieve a measurement accuracy within the sub-degree and millimeter level.
Another system consisting of two 2D laser scanners perpendicular to each other used four beam points, one on each edge of a rectangular object such as a container ship, to track the dynamic pose \cite{kim2014note}.
These studies used cameras and did not utilize LiDAR, but showed that 6-DOF pose of an object could be tracked by using only a few laser beam points.

Inspired by the aforementioned works, this paper proposes an algorithm for estimating the pose of a LiDAR frame with respect to the target frame using only four beam points, which are the closest points to each target corner. Unlike previous works, this paper uses only the 3D LiDAR for static pose estimation and proposes a method to extract the feature points and preprocess them to minimize the effect of high-level depth measurement noise.

\section{Sensor Misalignment Estimation}


\begin{figure*}[htb]
\centering
\includegraphics[width=0.9\textwidth]{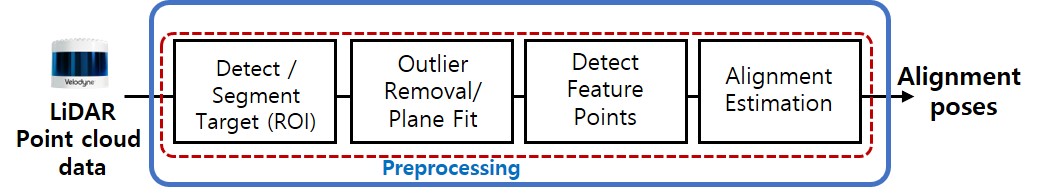}
\caption{\label{fig:overview_mod} Overview of Proposed Algorithm
}
\end{figure*}

The proposed algorithm can estimate the 6-DOF alignment, three orientations, and three positions of the LiDAR on a vehicle or robot . However, this paper is focused on evaluating the estimation accuracy of the orientations and the horizontal position of a LiDAR installation, which are critical alignments in terms of safety. The vertical position of a LiDAR can be varied depending on the vehicles, and the distance measurement of the target board is not necessary for calibration.
An overview for estimating the yaw and tilt angles is described in Fig. \ref{fig:overview_mod}.



\subsection{Preprocessing and Feature Point Extraction}

\begin{figure}
\centering
\includegraphics[width=0.48\textwidth]{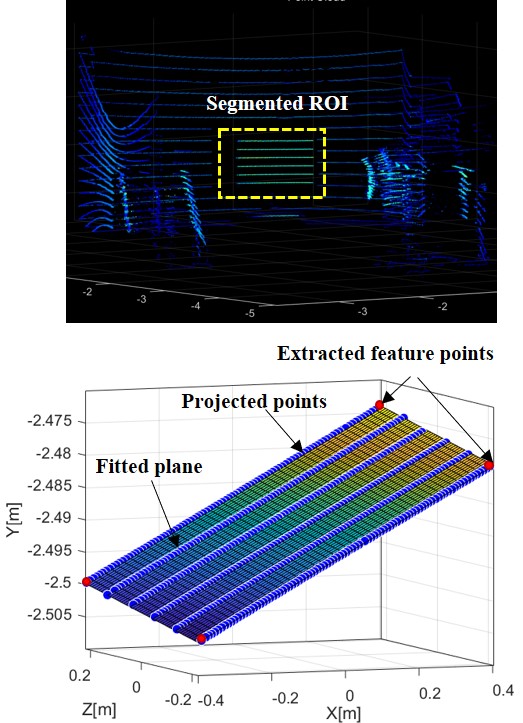}
\caption{ ROI segmented from one scan of LiDAR point cloud. The feature beam points are segmented from the beam points that are projected on the fitted plane}
\label{fig:lidarscan}
\end{figure}

One scan of a LiDAR gathers a vast number of 3D points. The preprocessing is conducted to segments the region of interest (ROI) of the target to reduce the overall computational load.
Since the flat surface of the target stands out clearly from the background data, as shown in Fig. \ref{fig:lidarscan}, the Euclidean clustering algorithm is applied to detect and segment the target surface automatically.
The clustering algorithm groups 3D points that have close Euclidean distances among one another. From the clustering candidates, the ROI is selected based on the conditions of the average distance, the number of points, and the average beam reflectivity values.

After segmenting the point cloud data on the target surface, the plane model of the target is estimated.
A plane model can be expressed by a normal vector, $\bf{n} = [a, b, c]^T$, and a distance $d$ such that for a point on the surface, $\bf{p} = [x, y, z]^T$.
The plane model can be estimated using the least-square method, but it may not give the optimal estimation because the point cloud data contains outliers and high depth noise (approximately 14 mm $1\sigma$). Thus, this study applied a random sample consensus (RANSAC)-based plane fitting algorithm, which can effectively remove outliers in the LiDAR point cloud.

The RANSAC plane fitting method is composed of two stages: hypothesis and verification. For the hypothesis, the method randomly selects 3 data points and estimates the plane normal vector. The distance error of all the data points from the estimated plane model is calculated. Then, it verifies the hypothesis by counting inliers that satisfy the threshold of the distance error. After some iterations, it determines the best-fit plane model that has the highest number of inlier data.

\subsection{Alignment Estimation}

\begin{figure}
\centering
\includegraphics[width=0.48\textwidth]{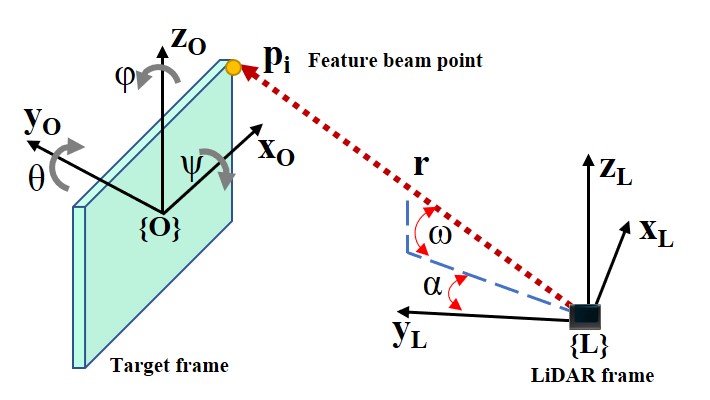}
\caption{\label{fig:coordinate} Coordinate frame of target board and the LiDAR}
\end{figure}

The coordinate frame of the target board and the LiDAR are defined in Fig. \ref{fig:coordinate}.
The positions of the beam points are measured in terms of azimuth ($\alpha$), vertical angle ($\omega$), and range ($r$). The azimuth and vertical angle are positive in the clockwise and upward directions, respectively. The position of the beam points can be converted to the Cartesian coordinate system of the LiDAR. ({L}) with the following equation:

\begin{equation}
 {}^{\bf{L}}{\bf{P}} =
    \left[ {\begin{array}{*{20}{c}}
    {{x_L}}\\
    {{y_L}}\\
    {{z_L}}
    \end{array}} \right] = r\left[ {\begin{array}{*{20}{c}}
    {\cos \omega \sin \alpha }\\
    {\cos \omega \cos \alpha }\\
    {\sin \alpha }
    \end{array}} \right]
    \label{eq_polar2cart}
\end{equation}
where the Z-axis is the vertical axis, the Y-axis is the distance axis, and the X-axis is the horizontal axis. Note that there can be some offset added to $Z_L$ depending on the relative position of each laser diode from the optical center.

The purpose of the proposed algorithm is to estimate the relative rotation and translation of the LiDAR sensor body with respect to the reference target board. Thus, the whole system is configured with two Cartesian coordinate frames of the target board (${O}$) and the LiDAR body (${L}$), as described in Fig. \ref{fig:coordinate}.

As shown in Fig.\ref{fig:coordinate}, the coordinate frame of the target (${O}$) is located at the geometric center of the board surface, and the coordinate frame ${L}$ is positioned at the optical center of the LiDAR.

The aim or yaw angle ($\theta$) is defined as the pointing angle with respect to the vertical axis (y-axis). The yaw misalignment generates a false relative horizontal (x-axis) and distance (y-axis) position of the obstacle. The tilt misalignment ($\phi$), which is the angle offset pointing upward or downward from the nominal orientation, causes a false vertical (z-axis) and distance position of the obstacle. The roll misalignment ($\psi$), which is the rotation with respect to the pointing direction (y-axis), can cause errors in the horizontal and vertical positions of an obstacle.

The 6-DOF poses of the LiDAR sensor from the target board can be estimated by solving for the transformation matrix ${}_L^OM(\phi, \theta, \psi, \Delta x,\Delta y,\Delta z)$. It describes the geometric relationship of the coordinate frame (${L}$) from the coordinate frame of (${O}$) and is an augmented matrix of the rotation matrix (${}_{\bf{L}}^{\bf{O}}{\bf{R}}$) and the translation vector (${}_{\bf{L}}^{\bf{O}}{\bf{T}}$). The rotation matrix is expressed in Euler angles with the convention of ${}_{\bf{L}}^{\bf{O}}{\bf{R}} = {R_z}(\phi){R_y}(\theta){R_x}(\psi)$.

The transformation matrix ${}_{\bf{L}}^{\bf{O}}{\bf{M}}$ can be solved by using the positions of four corner points (${}^{\bf{O}}{\bf{p_i}}$) and the corresponding LiDAR beam points (${}^{\bf{L}}{\bf{p_i}}$) with the relationship of

\begin{equation}
    {}^{\bf{O}}{\bf{P}} = {}_{\bf{L}}^{\bf{O}}{\bf{M}}{}^{\bf{L}}{\bf{P}} = [{}_{\bf{L}}^{\bf{O}}{\bf{R}}|{}_{\bf{L}}^{\bf{O}}{\bf{T}}]{}^{\bf{L}}{\bf{P}}
    \label{eq_mat}
\end{equation}.

The matrix (${}^{\bf{O}}{\bf{P }}$) is composed of the corner points of (${}^{\bf{O}}{\bf{p_i}}$) for i=1 to 4, which are known values in terms of the target dimension of W(mm) by H(mm) as

\begin{equation}
{}^{\bf{O}}{\bf{P}} = \left[ {\begin{array}{*{20}{c}}
{ - W/2}&{W/2}&{ - W/2}&{W/2}\\
0&0&0&0\\
{H/2}&{H/2}&{ - H/2}&{ - H/2}
\end{array}} \right]
\label{eq_OP}
\end{equation}.

\begin{figure}
\centering
\includegraphics[width=0.45\textwidth]{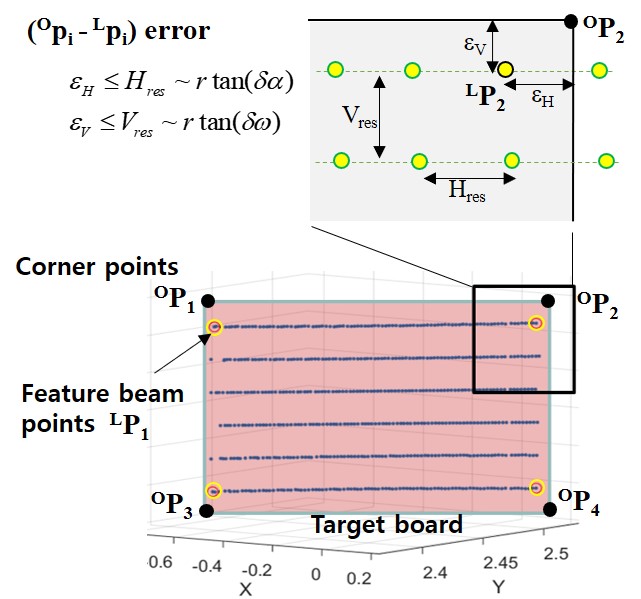}
\caption{\label{fig:targetboard} The position error between corners to feature beam points}
\end{figure}

The matrix (${}^{\bf{L}}{\bf{P }}$) is composed of the corresponding beam points that are the closest to each corner (${}^{\bf{L}}{\bf{p_i}}$) for i=1 to 4. Due to the low resolution of the LiDAR, they are not the exact match of the target corners and have the uncertainty bound shown in Fig. \ref{fig:targetboard}.

With the measurements of the feature beam points, the six pose variables can be estimated by minimizing the cost function $F(\beta)$, where $\beta$ is the 6-dimensional vector composed of translation and rotation components as
  
\begin{align*} 
{\bf{F}}(\beta ) = \frac{1}{2}\sum {\left\| {{}^{\bf{O}}{\bf{P}} - [{}_{\bf{L}}^{\bf{O}}{\bf{R}}|{}_{\bf{L}}^{\bf{O}}{\bf{T}}]{}^{\bf{L}}{\bf{P}}} \right\|} \\
\beta  = (\phi ,\theta ,\psi ,\Delta x,\Delta y,\Delta z)
\end{align*}

The cost function vector $F(\beta)$ is composed of four subfunctions, one for each feature point as ${\bf{F(\beta)}} = \left[ {{F_1};{F_2};{F_3};{F_4}} \right]$, where $F(\beta)_i$ is defined in Eq. \cite{eq_Fi}.

\begin{figure*}[!t]
\normalsize
\begin{equation}
F{(\beta )_i} = \left[ {\begin{array}{*{20}{c}}
{{}^O{P_{1,i}} + {r_i}(c{\alpha _i}c{\omega _i})(c\psi s\phi  - c\phi s\psi s\theta ) - {r_i}(s{\alpha _i})(s\psi s\phi  + c\psi c\phi s\theta ) - {r_i}(c{\alpha _i}s{\omega _i})(c\theta c\phi ) - \Delta x}\\
{{r_i}(s{\alpha _i})(c\phi s\psi  - c\psi s\theta s\phi ) - {r_i}(c{\alpha _i}c{\omega _i})(c\psi c\phi  + s\psi s\theta s\phi ) - {r_i}(c{\alpha _i}s{\omega _i})c\theta s\phi  - \Delta y}\\
{{}^O{P_{3,i}} + {r_i}(c{\alpha _i}s{\omega _i})s\theta  - {r_i}(s{\alpha _i})c\psi c\theta  - {r_i}(c{\alpha _i}c{\omega _i})c\theta s\psi  - \Delta z}
\end{array}} \right]
\label{eq_Fi}
\end{equation}
\hrulefill
\end{figure*}

The optimal solution of the pose variables, $\beta^*$, is the minimum point of the cost function obtained from applying a nonlinear least square optimization method. This paper used the Levenberg–Marquardt algorithm to solve for the optimal pose variables $\beta^*$ by iteratively estimating the optimal $\beta^*$.

\begin{equation}
{\beta ^*} = \arg {\min _\beta }{\bf{F}}(\beta )
\end{equation}

\begin{equation}
{\beta ^*} = \beta  - \eta {({{\bf{J}}^{\bf{T}}}{\bf{J}} + \lambda {\rm{diag}}({\bf{J}}))^{ - 1}}{{\bf{J}}^{\bf{T}}}{\bf{F}}(\beta ))
\end{equation}
where $\bf{J}$ is the Jacobian matrix for cost function $\bf{F}$, a constant $\eta$=0.02, and the varying tuning rate $\lambda$ is initially set at 0.3.

\section{Simulation}
The closeness of the feature beam points to the actual target corners is related to the vertical and azimuth angular resolutions of the LiDAR system. The lower the angular resolutions are, the higher the resulting alignment error.
Additionally, the noise and offset of depth measurement are other factors that cause alignment estimation errors, especially for yaw and tilt angles.
Therefore, in this section, simulations are conducted to analyze the level of precision and accuracy of the proposed algorithm with the given sensor specification (Velodyne VLP-16)

\subsection{Analysis of LiDAR resolution and noise}
The measurements of the feature beam points are contaminated by angular resolution uncertainty ($\delta\alpha \delta\omega $), depth measurement noise ($\epsilon$) and offset ($\Delta r$). Thus, the converted positions in the Cartesian coordinate also contain the measurement uncertainty as


\begin{equation}
^L{\widehat {\bf{p}}_i} =  ({r_i} + \varepsilon )\left[ {\begin{array}{*{20}{c}}
{\cos ({\alpha _i} + \delta \alpha )\sin ({\omega _i} + \delta \omega )}\\
{\cos ({\alpha _i} + \delta \alpha )\cos ({\omega _i} + \delta \omega )}\\
{\sin ({\alpha _i} + \delta \alpha )}
\end{array}} \right]
\end{equation}

According to the Velodyne VLP-16 datasheet, the vertical angle resolution ($\delta_{\omega}$) is 2 degrees, the azimuth angle resolution ($\delta_{\alpha}$) is 0.2 degrees, and the range measurement noise ($\epsilon \sim{\rm N}(\Delta r,{\sigma ^2}))$ is up to 14 mm ($1\sigma$) with an offset ($\Delta r$) up to 5 millimeters.
With this specification, the horizontal distance between the two consecutive beam points in the same vertical channel is approximately 9 mm, and the vertical gap between the two beam channels is approximately 90 mm when the target board is placed 2.5 meters from the sensor.

Additionally, the spinning speed fluctuation is within 3 rpm, which slightly varies the positions of the feature beams for each scan. The azimuth angle of a feature point beam varies within 0.2 degrees, which can result in an X-axis translation error as high as 9 mm.
A simple calculation shows that a depth measurement noise of 30 mm can cause up to 2 degrees of error in yaw angle estimation when the target is 1 meter wide and placed 2.5 meters away.
To minimize the effect of the high noise of LiDAR, this paper proposes preprocessing that projects the feature beams on the best-fit plane estimated from the point cloud data in the ROI region.

\subsection{Simulation result}
\begin{figure}
\centering
\includegraphics[width=0.48\textwidth]{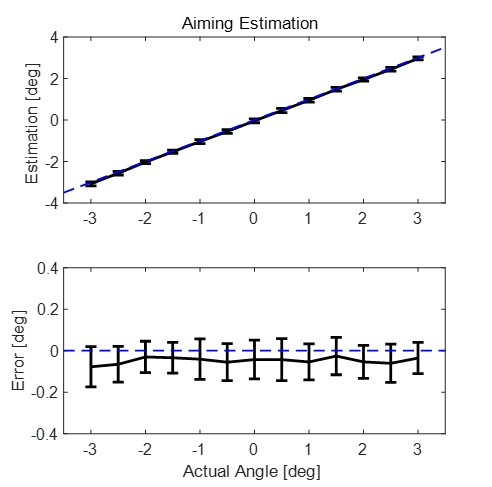}
\caption{\label{fig:sim_aim} Simulation for accuracy and precision of yaw angle estimation for reference angle from -3 to 3 degrees
}
\end{figure}

\begin{figure}
\centering
\includegraphics[width=0.48\textwidth]{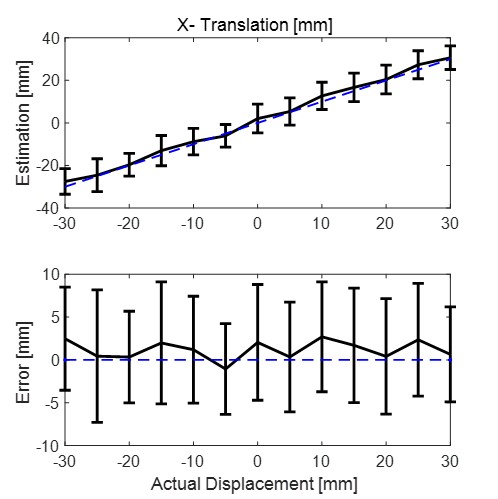}
\caption{\label{fig:sim_x} Simulation for accuracy and precision of X-axis position for reference displacement from -30 to 30 millimeters
}
\end{figure}

A Monte Carlo simulation is applied to evaluate the precision and accuracy of the proposed estimation algorithm based on the sensor specification. The measurement uncertainty due to the angular resolution and depth measurement is applied as described in the previous section.
The width and height of the target board are set to be 900 mm and 540 mm, respectively. The initial position of the target board is configured to be 2.5 meters and 0.7 meters in the y-axis and x-axis directions of the target board frame, respectively.
First, the yaw estimation is evaluated by giving a reference yaw angle from -3 to 3 degrees with a step of 0.5 degrees. As shown in Fig. \ref{fig:sim_aim}, the yaw estimation has an offset of approximately 0.05 degrees and a precision of 0.09 degrees.
A similar simulation is conducted for x-position estimation within a range of -30 to 30 millimeters. It shows that the estimation has accuracy within an average error of 1.2 mm and a precision of 6.4 mm, as shown in Fig. \ref{fig:sim_x}.

The final simulation is conducted with random orientations for all angles within 3 degrees, and horizontal displacement within 30 mm from the initial pose is configured. For each pose, 50 scans are used to calculate the mean and standard deviation for that specific pose. The overall accuracy and precision are then evaluated by averaging the results for each pose.

As summarized in Table \ref{tab:sim}, the proposed method has an estimation performance of precision and accuracy within 0.3 degrees and 5 mm in the configured motion bound.

The roll estimation shows the highest error among the orientations. This is because the roll error mainly depends on the position deviation of feature beam points from the corners, which are the X-axis and Z-axis positions of the beam points. If the LiDAR has a higher angular resolution that can reduce the uncertainty of feature beam positions, then the error in the roll estimation can be reduced.

Additionally, the tilt angle shows a lower precision than the yaw estimation due to the low vertical resolution of the LiDAR. The tilt estimation is dependent on the vertical position uncertainty of the feature beams from the actual corners. Thus, a higher vertical resolution of beams lowers the position uncertainty, which can improve the tilt estimation accuracy.

\begin{figure}
\centering
\includegraphics[width=0.48\textwidth]{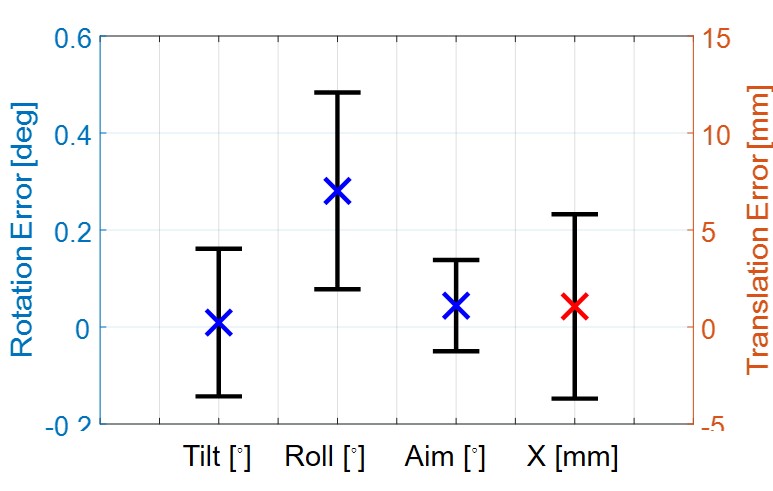}
\caption{\label{fig:sim_random} Simulation for accuracy and precision of random orientations and displacement}
\end{figure}

\begin{table}[t]
\caption{ Simulation result of Alignment Estimation}
\label{tab:sim}
\center
{\renewcommand{\arraystretch}{1.5}
   \begin{tabular*}{\linewidth}{c|cccc}
    \hline
    Estimation& Tilt $(^{\circ})$&	Roll $(^{\circ})$&Yaw $(^{\circ})$ &  $\Delta X (mm)$\\
    \hline
    Accuracy&  0.01&   0.28    &0.04   &1.1\\
    Precision&  0.15&   0.20&   0.1&   4.8\\
    \hline
    \end{tabular*}
}
\end{table}

\section{Experiment}

A test bench was designed to evaluate the accuracy and precision of the proposed alignment system.
The test bench is composed of two parts: a target board module and a sensor control module.

\begin{figure}
\centering
\includegraphics[width=0.48\textwidth]{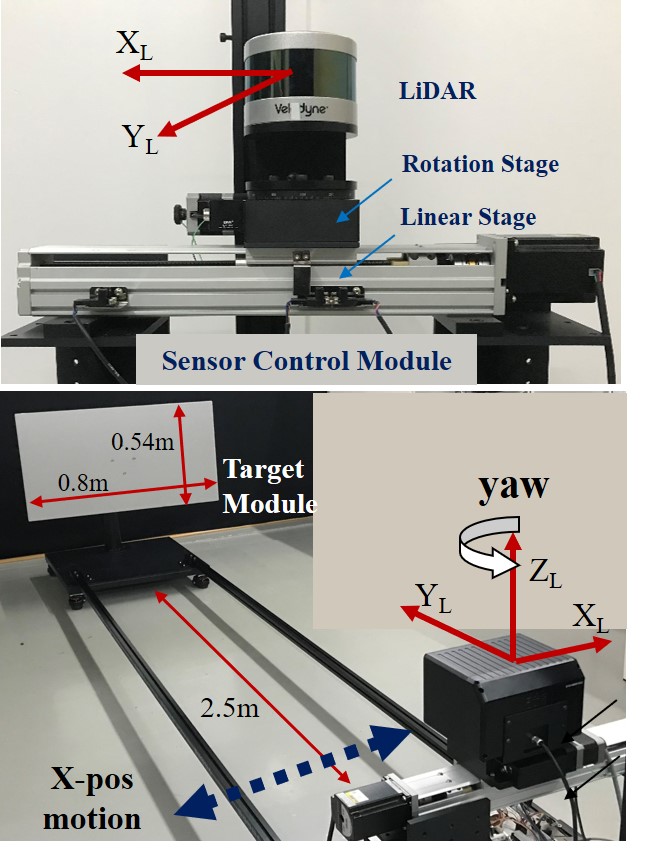}
\caption{\label{fig:expsetup} Experiment test bench with a LiDAR on rotation stage
}
\end{figure}

\subsection{Target board module}
The target board is built with a rigid flat aluminum board with dimensions of 0.9 m width and 0.54 m height.
The center of the board is positioned 0.5 m from the ground facing the sensor control module, which is placed at a 2.5 m distance.
The base for the target has four castor wheels that can adjust the hardware height on an uneven floor.

The coordinate frame of the target board ({O}) is set at the geometric center of the board with a positive X-axis to the right side and a negative Y-axis in the direction toward the sensor module

\subsubsection{Sensor Control Module}
The sensor control module is composed of the LiDAR, a linear motor stage, a rotation motor stage, and a motor controller.
The LiDAR is installed on the rotation stage, which is placed on top of the linear motor stage. The rotation stage can control the yaw angle of the LiDAR with a subdegree level of precision, and the linear stage can control the x-position of the sensor with an accuracy of 0.01 mm.
The reference yaw angle and the x-position of the LiDAR are controlled by the motor controller based on Arduino Mega.

The base of the sensor control module is firmly connected to the target module with two long aluminum profile beams such that the relative position between the two modules is 2.5 meters in the Y-direction and -0.7 meters in the X-direction.

\subsection{Experiment Results}
\begin{figure}
\centering
\includegraphics[width=0.48\textwidth]{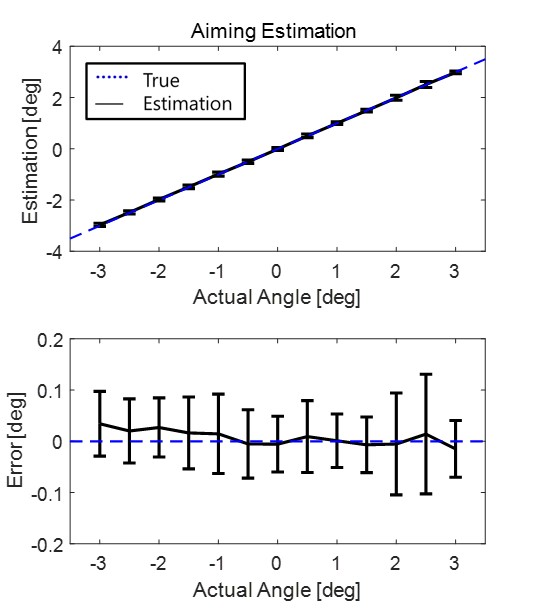}
\caption{\label{fig:aim_case11} Accuracy and precision of yaw angle estimation for reference angle from -10 to 10 degrees
}
\end{figure}

\begin{figure}
\centering
\includegraphics[width=0.48\textwidth]{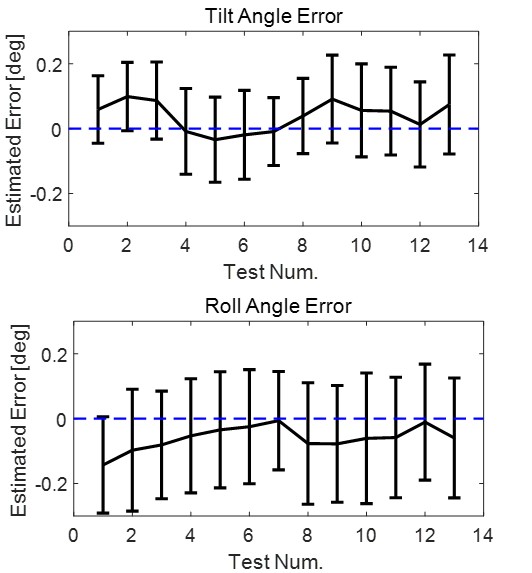}
\caption{\label{fig:tilt_roll_case11} Accuracy and precision of tilt and roll estimation for reference angle from -10 to 10 degrees
}
\end{figure}

For the first test, a reference yaw rotation of the LiDAR is controlled from -10 to 10 degrees, and the alignment angles are estimated from 100 scans.
Throughout the test, the tilt and roll of the LiDAR are set at a constant value because the motor stage is limited to rotating only in the yaw direction.
The estimation errors for each angle are plotted in Fig. \ref{fig:aim_case11} and \ref{fig:tilt_roll_case11}. The results show that the error mean and precision for yaw estimation is less than 0.1 degrees. The tilt and roll estimation show a higher error up to 0.2 degrees, as expected from the simulation.

\begin{figure}
\centering
\includegraphics[width=0.48\textwidth]{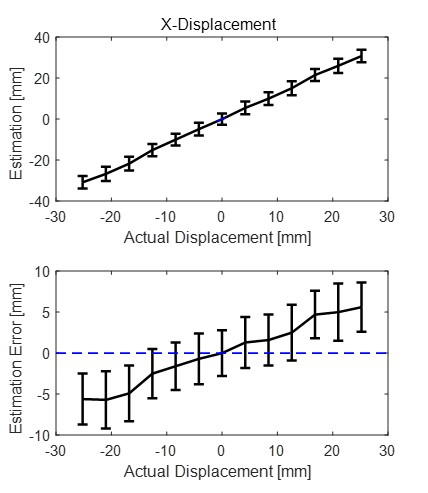}
\caption{\label{fig:x_case21} Accuracy and precision of X-position estimation
}
\end{figure}

\begin{figure}
\centering
\includegraphics[width=0.48\textwidth]{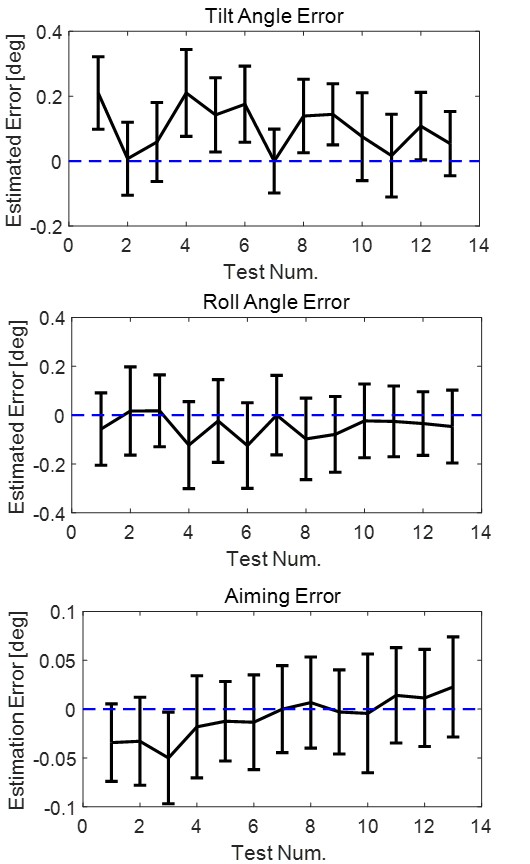}
\caption{\label{fig:angle_case21} Accuracy and precision of orientation for reference translation from -30 to 30 mm
}
\end{figure}

For the second test, only the x-position of LiDAR is varied from -30 mm to 30 mm from the initial position. As shown in Fig. \ref{fig:x_case21}, the algorithm is able to track the displacement with an error of less than 5 mm for accuracy and precision. While the position of the LiDAR is changed, the orientation estimation is maintained within 0.1 degrees for yaw and 0.2 degrees for tilt and roll, as shown in Fig. \ref{fig:angle_case21}

\begin{figure}
\centering
\includegraphics[width=0.48\textwidth]{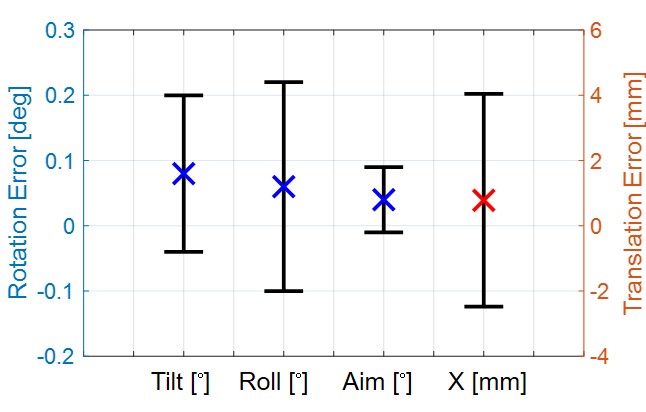}
\caption{\label{fig:expAll} Accuracy and precision of orientation for reference translation from -30 to 30 mm
}
\end{figure}

More tests were conducted by varying the yaw and horizontal displacement of LiDAR in combination. The overall accuracy and precision are summarized in Table \ref{tab:exp} and Fig. \ref{fig:expAll}.
The results confirm that the proposed algorithm can estimate the sensor alignment orientation with an average accuracy and precision of less than 0.2 degrees. Additionally, the horizontal position of the sensor can be estimated with an average precision of 3 mm.

Alignment estimation with subdegree accuracy has high performance compared to previous research and for most practical applications. If the LiDAR is replaced with a higher resolution model, then the proposed algorithm can provide a higher level of performance for sensor alignment inspection.

\begin{table}[t]
\caption{ Experiment Result of Alignment Estimation}
\label{tab:exp}
\center
{\renewcommand{\arraystretch}{1.5}
   \begin{tabular*}{\linewidth}{c|cccc}
    \hline
    Estimation& Tilt $(^{\circ})$&	Roll $(^{\circ})$&Yaw $(^{\circ})$ &  $\Delta X (mm)$\\
    \hline
    Accuracy&  0.08&   0.06    &0.03   &2.62\\
    Precision&  0.12&   0.16&   0.06&   3.5\\
    \hline
    \end{tabular*}
}
\end{table}

\section{Conclusion}
\label{conclusion}
This paper proposed a novel inspection system for estimating LiDAR misalignment after sensor attachment on a mobility body frame such as a vehicle, drone, or robot.
The proposed system uses only one target board at the fixed position and does not require any other sensor, such as a camera, which makes it simple to install and use for practical industrial applications.
The orientation and position of the LiDAR body with respect to the target board and the mobility body frame can be estimated by using four corners of the rectangular target board and the corresponding laser beam points. Using only a few points instead of the whole point cloud helps the algorithm estimate the sensor alignment in real time, requiring approximately 60 ms for one pose from a LiDAR scan.
Simulations and experimental results showed that the proposed system could estimate the alignment of a low-resolution 16 channel LiDAR with an error lower than 0.2 degrees for orientations and 4 mm for the horizontal position. The proposed algorithm shows a higher performance level if the applied LiDAR has a higher angular resolution.

However, due to the complexity of the test-bench design, this study was limited in controlling the reference yaw and horizontal position for the experiment.
Therefore, an advanced test bench with a 3-DOF rotation motor stage should be designed to investigate each orientation of the LiDAR alignment in future work.
Additionally, the algorithm should be improved to reduce the estimation error by applying more advanced filters and combinations with other sensors.


\bibliographystyle{IEEEtran}
\bibliography{arxiv_manuscript_submit}

\begin{thebibliography}{10}
\providecommand{\url}[1]{#1}
\csname url@samestyle\endcsname
\providecommand{\newblock}{\relax}
\providecommand{\bibinfo}[2]{#2}
\providecommand{\BIBentrySTDinterwordspacing}{\spaceskip=0pt\relax}
\providecommand{\BIBentryALTinterwordstretchfactor}{4}
\providecommand{\BIBentryALTinterwordspacing}{\spaceskip=\fontdimen2\font plus
\BIBentryALTinterwordstretchfactor\fontdimen3\font minus
  \fontdimen4\font\relax}
\providecommand{\BIBforeignlanguage}[2]{{%
\expandafter\ifx\csname l@#1\endcsname\relax
\typeout{** WARNING: IEEEtran.bst: No hyphenation pattern has been}%
\typeout{** loaded for the language `#1'. Using the pattern for}%
\typeout{** the default language instead.}%
\else
\language=\csname l@#1\endcsname
\fi
#2}}
\providecommand{\BIBdecl}{\relax}
\BIBdecl

\bibitem{fremont2008extrinsic}
V.~Fremont, P.~Bonnifait \emph{et~al.}, ``Extrinsic calibration between a
  multi-layer lidar and a camera,'' in \emph{2008 IEEE International Conference
  on Multisensor Fusion and Integration for Intelligent Systems}.\hskip 1em
  plus 0.5em minus 0.4em\relax IEEE, 2008, pp. 214--219.

\bibitem{alismail2012automatic}
H.~Alismail, L.~D. Baker, and B.~Browning, ``Automatic calibration of a range
  sensor and camera system,'' in \emph{2012 Second International Conference on
  3D Imaging, Modeling, Processing, Visualization \& Transmission}.\hskip 1em
  plus 0.5em minus 0.4em\relax IEEE, 2012, pp. 286--292.

\bibitem{geiger2012automatic}
A.~Geiger, F.~Moosmann, {\"O}.~Car, and B.~Schuster, ``Automatic camera and
  range sensor calibration using a single shot,'' in \emph{2012 IEEE
  International Conference on Robotics and Automation}.\hskip 1em plus 0.5em
  minus 0.4em\relax IEEE, 2012, pp. 3936--3943.

\bibitem{pusztai2017accurate}
Z.~Pusztai and L.~Hajder, ``Accurate calibration of lidar-camera systems using
  ordinary boxes,'' in \emph{Proceedings of the IEEE International Conference
  on Computer Vision Workshops}, 2017, pp. 394--402.

\bibitem{moosmann2011velodyne}
F.~Moosmann and C.~Stiller, ``Velodyne slam,'' in \emph{2011 IEEE Intelligent
  Vehicles Symposium (IV)}.\hskip 1em plus 0.5em minus 0.4em\relax IEEE, 2011,
  pp. 393--398.

\bibitem{segal2009generalized}
A.~Segal, D.~Haehnel, and S.~Thrun, ``Generalized-icp.'' in \emph{Robotics:
  science and systems}, vol.~2, no.~4.\hskip 1em plus 0.5em minus 0.4em\relax
  Seattle, WA, 2009, p. 435.

\bibitem{muhammad2010calibration}
N.~Muhammad and S.~Lacroix, ``Calibration of a rotating multi-beam lidar,'' in
  \emph{2010 IEEE/RSJ International Conference on Intelligent Robots and
  Systems}.\hskip 1em plus 0.5em minus 0.4em\relax IEEE, 2010, pp. 5648--5653.

\bibitem{bergelt2017improving}
R.~Bergelt, O.~Khan, and W.~Hardt, ``Improving the intrinsic calibration of a
  velodyne lidar sensor,'' in \emph{2017 IEEE SENSORS}.\hskip 1em plus 0.5em
  minus 0.4em\relax IEEE, 2017, pp. 1--3.

\bibitem{kim2019extrinsic}
E.-S. Kim and S.-Y. Park, ``Extrinsic calibration of a camera-lidar multi
  sensor system using a planar chessboard,'' in \emph{2019 Eleventh
  International Conference on Ubiquitous and Future Networks (ICUFN)}.\hskip
  1em plus 0.5em minus 0.4em\relax IEEE, 2019, pp. 89--91.

\bibitem{zhang2004extrinsic}
Q.~Zhang and R.~Pless, ``Extrinsic calibration of a camera and laser range
  finder (improves camera calibration),'' in \emph{2004 IEEE/RSJ International
  Conference on Intelligent Robots and Systems (IROS)(IEEE Cat. No.
  04CH37566)}, vol.~3.\hskip 1em plus 0.5em minus 0.4em\relax IEEE, 2004, pp.
  2301--2306.

\bibitem{verma2019automatic}
S.~Verma, J.~S. Berrio, S.~Worrall, and E.~Nebot, ``Automatic extrinsic
  calibration between a camera and a 3d lidar using 3d point and plane
  correspondences,'' in \emph{2019 IEEE Intelligent Transportation Systems
  Conference (ITSC)}.\hskip 1em plus 0.5em minus 0.4em\relax IEEE, 2019, pp.
  3906--3912.

\bibitem{zhou2018automatic}
L.~Zhou, Z.~Li, and M.~Kaess, ``Automatic extrinsic calibration of a camera and
  a 3d lidar using line and plane correspondences,'' in \emph{2018 IEEE/RSJ
  International Conference on Intelligent Robots and Systems (IROS)}.\hskip 1em
  plus 0.5em minus 0.4em\relax IEEE, 2018, pp. 5562--5569.

\bibitem{gong20133d}
X.~Gong, Y.~Lin, and J.~Liu, ``3d lidar-camera extrinsic calibration using an
  arbitrary trihedron,'' \emph{Sensors}, vol.~13, no.~2, pp. 1902--1918, 2013.

\bibitem{wang2014simulation}
F.~Wang, ``Simulation of registration accuracy of iterative closest point (icp)
  method for pose estimation,'' in \emph{Applied Mechanics and Materials}, vol.
  475.\hskip 1em plus 0.5em minus 0.4em\relax Trans Tech Publ, 2014, pp.
  401--404.

\bibitem{heide2018calibration}
N.~Heide, T.~Emter, and J.~Petereit, ``Calibration of multiple 3d lidar sensors
  to a common vehicle frame,'' in \emph{ISR 2018; 50th International Symposium
  on Robotics}.\hskip 1em plus 0.5em minus 0.4em\relax VDE, 2018, pp. 1--8.

\bibitem{zaganidis2017semantic}
A.~Zaganidis, M.~Magnusson, T.~Duckett, and G.~Cielniak, ``Semantic-assisted 3d
  normal distributions transform for scan registration in environments with
  limited structure,'' in \emph{2017 IEEE/RSJ International Conference on
  Intelligent Robots and Systems (IROS)}.\hskip 1em plus 0.5em minus
  0.4em\relax IEEE, 2017, pp. 4064--4069.

\bibitem{choi2012fast}
W.-S. Choi, Y.-S. Kim, S.-Y. Oh, and J.~Lee, ``Fast iterative closest point
  framework for 3d lidar data in intelligent vehicle,'' in \emph{2012 IEEE
  Intelligent Vehicles Symposium}.\hskip 1em plus 0.5em minus 0.4em\relax IEEE,
  2012, pp. 1029--1034.

\bibitem{wang2018single}
Y.-T. Wang, C.-C. Peng, A.~A. Ravankar, and A.~Ravankar, ``A single lidar-based
  feature fusion indoor localization algorithm,'' \emph{Sensors}, vol.~18,
  no.~4, p. 1294, 2018.

\bibitem{kim2012developing}
Y.-K. Kim, Y.~Kim, Y.~S. Jung, I.~G. Jang, K.-S. Kim, S.~Kim, and B.~M. Kwak,
  ``Developing accurate long-distance 6-dof motion detection with
  one-dimensional laser sensors: Three-beam detection system,'' \emph{IEEE
  Transactions on Industrial Electronics}, vol.~60, no.~8, pp. 3386--3395,
  2012.

\bibitem{kim2014design}
Y.-K. Kim, I.~G. Jang, K.-S. Kim, and S.~Kim, ``Design improvement of the
  three-beam detector towards a precise long-range 6-degree of freedom motion
  sensor system,'' \emph{Review of Scientific Instruments}, vol.~85, no.~1, p.
  015004, 2014.

\bibitem{kim2015portable}
Y.-K. Kim, K.-S. Kim, and S.~Kim, ``A portable and remote 6-dof pose sensor
  system with a long measurement range based on 1-d laser sensors,'' \emph{IEEE
  Transactions on Industrial Electronics}, vol.~62, no.~9, pp. 5722--5729,
  2015.

\bibitem{kim2014note}
Y.-K. Kim and K.-S. Kim, ``Note: Reliable and non-contact 6d motion tracking
  system based on 2d laser scanners for cargo transportation,'' \emph{Review of
  Scientific Instruments}, vol.~85, no.~10, p. 106102, 2014.

\end{thebibliography}

\end{document}